\documentclass[preprint,12pt, a4paper]{elsarticle}

\usepackage{amsmath}
\usepackage{amssymb}
\usepackage{graphicx}
\usepackage{float}
\usepackage{booktabs}
\usepackage{epstopdf}
\usepackage[margin=2.5cm]{geometry}
\usepackage{multicol}
\usepackage{pdflscape}
\usepackage{array}
\usepackage{xcolor}
\usepackage{tablefootnote}
\usepackage{amsthm}
\usepackage{mathtools}

\usepackage{lineno}

\usepackage{multirow}
\usepackage{tikz}
\usetikzlibrary{shapes.geometric, arrows}
\tikzstyle{startstop} = [rectangle, rounded corners, minimum width=2cm, minimum height=0.75cm,text centered, draw=black, fill=red!30]
\tikzstyle{io} = [trapezium, trapezium left angle=70, trapezium right angle=110, minimum width=2cm, minimum height=0.75cm, text centered, draw=black, fill=blue!30]
\tikzstyle{process} = [rectangle, minimum width=2cm, minimum height=0.75cm, text centered, draw=black, fill=orange!30]
\tikzstyle{decision} = [diamond, minimum width=2cm, minimum height=0.75cm, text centered, draw=black, fill=green!30]
\tikzstyle{arrow} = [thick,->,>=stealth]

\journal{SoftwareX}
\begin{document}

\begin{frontmatter}



\title{OpenQSEI: a MATLAB package for Quasi Static Elasticity Imaging}


\author[label1]{Danny Smyl}
\author[label1]{Sven Bossuyt}
\author[label2,label3,label4]{Dong Liu}

\address[label1]{Department of Mechanical Engineering, Aalto University, Espoo, Finland}
\address[label2]{CAS Key Laboratory of Microscale Magnetic Resonance and Department of Modern Physics, University of Science and Technology of China (USTC), Hefei 230026, China}
\address[label3]{Hefei National Laboratory for Physical Sciences at the Microscale, USTC, China}
\address[label4]{Synergetic Innovation Center of Quantum Information and Quantum Physics, USTC, China}

\begin{abstract}
Quasi Static Elasticity Imaging (QSEI) aims to computationally reconstruct the inhomogeneous distribution of the elastic modulus using a measured displacement field.
QSEI is a well-established imaging modality used in medical imaging for localizing tissue abnormalities.
More recently, QSEI has shown promise in applications of structural health monitoring and materials characterization.
Despite the growing usage of QSEI in multiple fields, fully-executable open source packages are not readily available.
To address this, OpenQSEI is developed using a MATLAB platform and shared in an online community setting for continual algorithmic improvement.
In this article, we describe the mathematical background of QSEI and demonstrate the basic functionalities of OpenQSEI with examples.

\end{abstract}


\end{frontmatter}


\section{Motivation, significance, and background}
\label{}
Quasi Static Elasticity Imaging (QSEI) is a well-established imaging modality used in medical imaging \cite{goenezen2011,gokhale2008,konofagou2004quo,oberai2003,barbone2002,parker1996techniques,kallel1996,ophir1991} and an increasingly popular modality in applications of structural health monitoring and materials characterization \cite{smyl2018,hoerig2017,bonnet2005inverse}.
Potential applications of QSEI include medical imaging of arteriosclerosis \cite{barbone2004elastic} or tissue abnormalities manifested by, for example, breast cancer \cite{konofagou2004quo}.
Some engineering applications of QSEI include detection of distributed or localized damage \cite{smyl2018} and characterization of unknown elastic parameters \cite{bonnet2005inverse}.
Owing to the multiplicity of QSEI applications, researchers have made significant progress in developing QSEI algorithms.
Recent improvements have, for example, included speedups using gradient-based approaches \cite{oberai2003}, inclusion of non-linear material models \cite{goenezen2011,gokhale2008}, and implementation of machine learning \cite{hoerig2017}.
Despite these developments, fully-executable open source QSEI software is currently unavailable.
To address this, we have developed an open source, MATLAB executable QSEI platform: OpenQSEI.
In the following we will provide mathematical background for QSEI/OpenQSEI and briefly discuss preliminaries of the mathematical functionalities and their software integration.

The aim of QSEI is to reconstruct distribution of the elasticity modulus $E$ using a measured displacement field $u_m$  with a computational inverse approach.
The observation model for such an inverse problem has the form 

\begin{equation}
\label{ob}
u_m = U(E) + e
\end{equation}

\noindent where $e$ is the assumed Gaussian-distributed noise and $U(E)$ is the forward model solved using a discretized numerical method (the FEM is adopted herein).
In a medical setting, $u_m$ is often provided by either (i) tracking the displacements of speckles using ultrasound or (ii) using phase information from the measured magnetic field obtained through Nuclear Magnetic Resonance Imaging. 
In structural applications, $u_m$ may be obtained using Digital Image Correlation or discrete measurements, such as strain gauges or extensometers.
For any field of application, having the ability to simulate $u_m$ with added noise is an excellent way of testing algorithmic performance, most often this is done using the forward model.

Based on the observation model for Eq. \ref{ob}, we may write the constrained and regularized Least-Squares (LS) solution as

\begin{equation}
\label{sol}
\ell = \mathrm{arg}\underset{\begin{subarray}{c} 
E_\mathrm{L} < E < E_\mathrm{U}\\
\end{subarray}}{\min}
\{||L_e(u_m - U(E))||^2 + p_E(E)\}
\end{equation}

\noindent where $p_E(E)$ is the regularization functional of choice, $L_e^T L_e = C_e^{-1}$ where $C_e$ is the
observation noise covariance matrix, $E_\mathrm{L}$ and $E_\mathrm{U}$ are the user-defined constraints, and $||\cdot||$ denotes the Euclidean norm.
The regularization term is included due to the ill-posed nature of the inverse problem, meaning that standard LS approaches may yield non-unique solutions.
The choice of the appropriate regularization functional (prior model) is important in obtaining accurate reconstructions.
For example, in cases where sharp features are present, Total Variation (TV) regularization may be most appropriate; however, in cases where $E$ is relatively smooth, smoothness promoting regularization ($L_2$ and weighted) may provide best results \cite{mueller2012}.

To solve the inverse problem, an iterative constrained Gauss-Newton (GN) optimization regime is used.
The regime is equipped with a line-search algorithm to determine the step size $s_k$ in the solution $\theta_k = \theta_{k-1} + s_k\bar{\theta}$ where $\theta_k$  is the current estimate and $\bar{\theta}$ is the LS update.
Such an approach requires the Jacobian $J=\frac{\partial U(E)}{\partial E}$ at each iteration $k$, which is computed using the perturbation method with the choice of first- or second-order central differencing.

The overall goals of this work are to (a) develop a user-friendly software package where the mathematics are condensed into easily executable and readable MATLAB functions and (b)  share the software in an online community setting for continual algorithmic improvement.
Based on the mathematics outlined herein and goal (a), we designed OpenQSEI to have the following basic MATLAB functions for each \emph{mathematical functionality}.  \emph{Forward model/data simulation:} Meshing, inclusion generation, and FEM solution to the elasticity problem (plane-stress).  \emph{Prior models:} $L_2$ smoothness, weighted smoothness, and TV. \emph{Constrained optimization:} Cost function (cf. Eq. \ref{sol}), barrier functions for constraints, generation of gradients/Hessians, and line search.
In addition, functionalities for plotting and visualizing are included.

%
%
%
%

\section{Software description}
\label{}

\subsection{Software overview}
\label{SA}

OpenQSEI is designed to allow the user to easily access different prior models and optimization functionalities for imaging the elastic modulus using experimental or simulated displacement fields.
To do this, the OpenQSEI environment includes the essential functions for (i) simulation of data and a FEM implementation of the forward model (ii) three commonly used prior models (TV, $L_2$, and weighted smoothness)\footnote{For a comprehensive description of TV, $L_2$, and weighted smoothness regularization techniques, we refer the reader to \cite{mueller2012,strong2003edge,kaipio1999inverse}.}, (iii) estimation of the optimization starting point (best homogeneous estimate), constraints using cubic polynomial barrier functions, computation of the Jacobian, and line search used during optimization and (iv) determination of gradients and Hessians related to the prior models and constraints.
A flow chart outlining the basic software architecture is highlighted in Fig. 1.

\begin{figure}[H]
\label{PR}
\centering

\begin{tikzpicture}[node distance=1.1cm]

\node (start) [startstop] {Start OpenQSEI};
\node (pro0) [process, below of=start] {Generate meshing using \texttt{meshgen}};
\node (in1) [io, below of=pro0] {Select prior model, problem constraints, and Jacobian type};
\node (pro1) [process, below of=in1] {Simulate data using \texttt{SimulateData}};
\node (pro2) [process, below of=pro1] {Add noise to data and generate $C_e$};
\node (pro3) [process, below of=pro2] {Determine starting point and best homogeneous estimate using \texttt{BestHomogeneousE}};
\node (pro4) [process, below of=pro3] {Generate prior matricies using either \texttt{WeightedSmoothnessPrior} or \texttt{getTVMat}};
\node (pro5) [process, below of=pro4] {Initialize GN optimization parameters using \texttt{costfun} and \texttt{gradshessians}};
\node (startGN) [startstop, below of=pro5] {GN optimization (while $\frac{\ell_k - \ell_{k-1}}{\ell_{k-1}} > $ tol)};
\node (pro6) [process, below of=startGN] {Compute $J$ using \texttt{PertubedJ}};
\node (pro7) [process, below of=pro6] {Compute LS update $\bar{\theta}$, $s_k$ using \texttt{linesearch\_uniform}, and $\ell_k$ using \texttt{costfun}};
\node (pro8) [process, below of=pro7] {Update $\ell_k$ using \texttt{costfun} and the gradients/Hessians using \texttt{gradshessians}};
\node (pro9) [process, below of=pro8] {Plot results};
\node (endGN) [startstop, below of=pro9] {End};

\draw [arrow] (start) -- (pro0);
\draw [arrow] (pro0) -- (in1);
\draw [arrow] (in1) -- (pro1);
\draw [arrow] (pro1) -- node[anchor=west] {} (pro2);
\draw [arrow] (pro2) -- (pro3);
\draw [arrow] (pro3) -- (pro4);
\draw [arrow] (pro4) -- (pro5);
\draw [arrow] (pro5) -- (startGN);
\draw [arrow] (startGN) -- (pro6);
\draw [arrow] (pro6) -- (pro7);
\draw [arrow] (pro7) -- (pro8);
\draw [arrow] (pro8) -- (pro9);
\draw [arrow] (pro9) -- (endGN);

\end{tikzpicture}
  \caption{Outline of the OpenQSEI architecture and primary functions.  Note that in the functions \texttt{costfun} and \texttt{PertubedJ} require evaluation of the forward model \texttt{FMDL}.}\label{outline}
\end{figure}
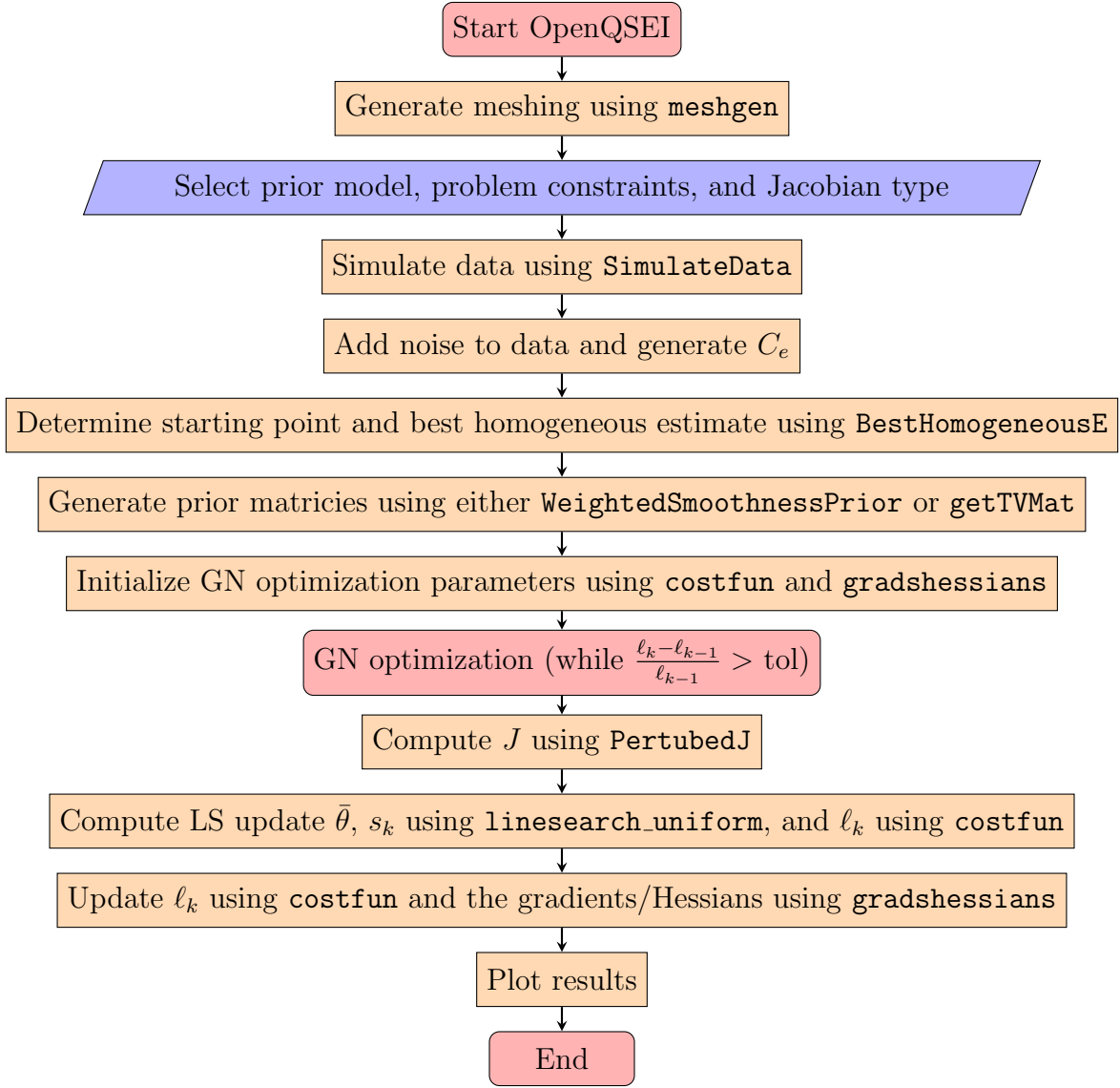

\subsection{Software functionalities}
\label{SF}
The primary software functionalities of OpenQSEI aim to solve the forward problem, provide appropriate prior models, and optimize the QSEI inverse problem.
For the forward in-plane elasticity problem, the package is equipped with a  FEM regime, \texttt{FMDL}.
\texttt{FMDL} uses piece-wise linear triangular elements, which are discretized utilizing user input geometry by employing MATLAB Delaunay Triangulation via \texttt{meshgen}.

Prior models using weighted smoothness and TV are provided through the implementation of functions \texttt{WeightedSmoothnessPrior} or \texttt{getTVMat}.
For \texttt{WeightedSmoothnessPrior}, the user may input spatial correlation length and the predicted range of $E$.
When employing \texttt{getTVMat}, the user may tailor the TV stabilization parameter $\beta$ and weighting parameter $\alpha$ for the physics of his/her particular problem.
The $L_2$ prior, on the other hand, is simply generated as a multiple of the identity matrix \textbf{I}, i.e. $\lambda \textbf{I}$.
Therefore only user input of the scalar $\lambda$ and generation of \textbf{I} are required during the optimization regime.

\begin{figure}[H]
  \centering
  \includegraphics[width=6.3in]{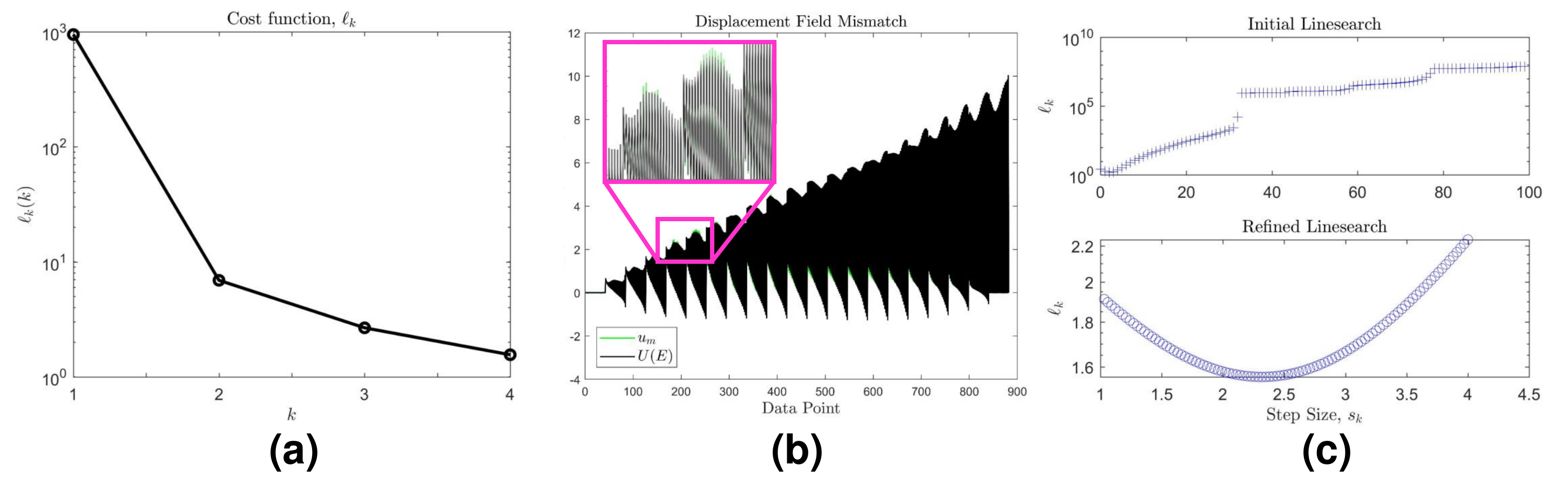}\\
  \caption{Figures generated during the optimization portion of OpenQSEI, (a) minimization of the cost function $\ell_k(k)$, (b) data mismatch of the measured $u_m$ and simulated displacement field $U(E)$, and (c) visualization of the line search using \texttt{linesearch\_uniform}.}\label{LSopt}
\end{figure}

To solve the inverse problem, the main MATLAB file \texttt{OpenQSEI} contains the shell LS-GN iterative optimization regime calling the necessary optimization functions.
The optimization regime is executed after initialization of all required parameters, such as meshing, initial guess via \texttt{BestHomogeneousE}, prior matricies, etc.
The optimization, which runs until the residual $\frac{\ell_k - \ell_{k-1}}{\ell_{k-1}} $ is below some user-defined tolerance (tol), includes iterative updating of the cost function, step size, and appropriate gradients/Hessians via \texttt{costfun}, \texttt{linesearch\_uniform}, and \texttt{gradshessians}.
Throughout the optimization process, the user is updated with useful figures including the minimization of the cost function, step-size calculation, and data mismatch.
A representative snippet of the optimization plots are shown in Fig. \ref{LSopt} after four iterations using the $L_2$ prior.

\section{Illustrative example - reconstruction of a unicorn}
\label{IE}
In the following, we provide an example of a unicorn-shaped elastic inclusion in a stretched plate.
The user interface is entirely included within the main MATLAB file \texttt{OpenQSEI.m}.
For this exercise, pre-loaded boundary conditions, external forces, and the image of a unicorn superimposed onto the meshing are contained in the files \texttt{cond.mat}, \texttt{force.mat}, and \texttt{Eunicorn.mat}, respectively.
Once the user has opened the file, he/she must change the directory to the OpenQSEI folder on their computing system.
After this preliminary, OpenQSEI is fully operable and the user may select the choice of prior model and constraints and execute the code.

\begin{figure}[H]
  \centering
  \includegraphics[width=6.25in]{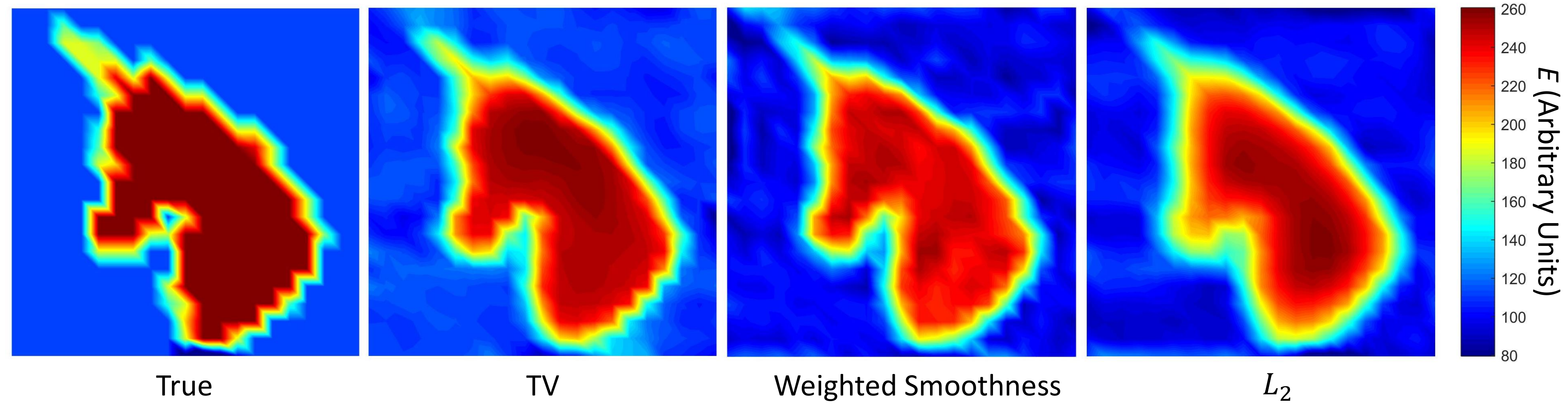}\\
  \caption{Reconstructions using OpenQSEI with three different prior models (labeled below each figure).}\label{Unicorn}
\end{figure}

In this example we reconstruct the true image of the unicorn, with 1.0\% noise standard deviation added to the simulated displacement field, using all three prior models (TV, weighted smoothness, and $L_2$ regularization).
We note that no attempt was made to optimize the regularization parameters, however the user may do so at his/her own discretion.
The constraints selected were $E_\mathrm{L}  = 0 < E < E_\mathrm{U}=300$ and the range of $E$ in the true unicorn image was $80 \leq E \leq 260$, where the units of $E$ are arbitrary.
Moreover, the stopping criteria during optimization was selected as $\frac{\ell_k - \ell_{k-1}}{\ell_{k-1}} \leq 10^{-6}$.
Reconstructions for this example are shown in Fig. \ref{Unicorn}.

Fig. \ref{Unicorn} demonstrates that all three priors resulted in satisfactory estimation of $E$ and captured the overall shape of the unicorn.
It should be remarked that the true image has sharp edges, which is best suited for TV prior and is visible in the overall improved reconstruction quality relative to the smoothness priors.
In cases where the unicorn or other distributions of $E$ are more smoothly dispersed, $L_2$ and weighted smoothness priors may be more appropriate prior models.

\section{Impact}
\label{}
Quasi-Static Elasticity Imaging (QSEI) is a powerful imaging tool used in various medical and engineering applications.
In medical imaging, QSEI is used for diagnosing tissue abnormalities and serious medical conditions, such as breast cancer and arteriosclerosis \cite{konofagou2004quo,barbone2004elastic}.
Potential applications in structural engineering include structural heath monitoring, damage detection, and material characterization \cite{smyl2018}.

Despite the clear relevance of QSEI in multiple research facets, to the best of our knowledge, there are no open-source QSEI packages available.
To that end, OpenQSEI is developed using a MATLAB-based platform to solve the non-linear ill-posed problem in a least-squares framework.
OpenQSEI users have the flexibility to select, experiment, and solve QSEI problems with different prior models, constraints, optimization parameters, geometries, boundary conditions, force fields, and more.

Due to the rapid advances with the field of QSEI over the recent years by numerous researchers (for example, \cite{hoerig2017, goenezen2011,gokhale2008}), we have decided to make OpenQSEI a community project, which will allow users to continually improve OpenQSEI via the online platform (see section \ref{LOC} for additional details).
Potential near-term advances improving the applicability and performance of OpenQSEI over the coming years may include (i) a faster semi-analytical regime for determining the Jacobian, (ii) higher-order $h,p,k$ forward models \cite{surana2016}, (iii) parallelization of the forward model for large-scale problems, (iv) additional prior models, for example $L_1$,  Lasso, and Ridge Regression methods, (v) estimation of unknown boundaries and/or forces, (vi) non-Gaussian noise models, and (vi) gradient-based methods \cite{oberai2003}.

\section{Conclusions}
\label{}
In this article, we presented the MATLAB-based Quasi Static Elasticity Imaging (QSEI) package: OpenQSEI.
The software promotes the open use of QSEI for users in a variety of fields, including medical imaging a structural engineering.
The present version of the software allows users the ability to solve QSEI problems with different prior models, constraints, optimization parameters, geometries, boundary conditions, force fields, and more.
In the future, via the OpenQSEI online platform, we hope to engage other researchers to further develop and improve the capabilities of OpenQSEI for a large suite of potential applications.

\section*{Acknowledgements}
\label{}
Authors DS and SB would like to acknowledge the support of the Department of Mechanical Engineering at Aalto University throughout this project.  
DL was supported by Anhui Provincial Natural Science Foundation (1708085MA25).
We would like to thank Chris Wolfe for his assistance in establishing the github repository.






%
%
%

\bibliographystyle{elsarticle-num} 
\bibliography{bib}

\section*{Required Metadata}
\label{LOC}
%

\begin{table}[!h]
\begin{tabular}{|l|p{6.5cm}|p{6.5cm}|}
\hline
\textbf{Nr.} & \textbf{Code metadata description} & \textbf{Please fill in this column} \\
\hline
C1 & Current code version & v1.0.2\\
\hline
C2 & Permanent link to code/repository used for this code version & \emph{https://github.com/openQSEI} \\
\hline
C3 & Legal Code License   & MIT \\
\hline
C4 & Code versioning system used & git \\
\hline
C5 & Software code languages, tools, and services used & MATLAB \\
\hline
C6 & Compilation requirements, operating environments \& dependencies & MATLAB 2017b or Octave 4.2.1 and either Mac OS X, Linux, or Windows\\
\hline
C7 & If available Link to developer documentation/manual &   \\
\hline
C8 & Support email for questions & danny.smyl@aalto.fi \\
\hline
\end{tabular}
\caption{Code metadata}
\label{} 
\end{table}

%


\end{document}